\documentclass[aps,prb,twocolumn,showpacs,floatfix,eqsecnum]{revtex4}
\usepackage{graphicx}
\usepackage{amssymb}
\usepackage{amsmath}
\usepackage{bm}

\begin{document}

\title{Flat-lens focusing of electrons on the surface of a topological
insulator}

\author{F. Hassler}
\affiliation{Instituut-Lorentz, Universiteit Leiden, P.O. Box 9506, 2300 RA
Leiden, The Netherlands}

\author{A. R. Akhmerov}
\affiliation{Instituut-Lorentz, Universiteit Leiden, P.O. Box 9506, 2300 RA
Leiden, The Netherlands}

\author{C. W. J. Beenakker}
\affiliation{Instituut-Lorentz, Universiteit Leiden, P.O. Box 9506, 2300 RA
Leiden, The Netherlands}

\date{May, 2010}

\begin{abstract}
  We propose the implementation of an electronic Veselago lens on the
  conducting surface of a three-dimensional topological insulator (such as
  Bi$_2$Te$_3$). The negative refraction needed for such a flat lens results
  from the sign change of the curvature of the Fermi surface, changing
  from a circular to a snowflake-like shape across a sufficiently large
  electrostatic potential step. No interband transition (as in graphene) is
  needed. For this reason, and because the topological insulator provides
  protection against backscattering, the potential step is able to focus
  a broad range of incident angles. We calculate the quantum interference
  pattern produced by a point source, generalizing the analogous optical
  calculation to include the effect of a noncircular Fermi surface (having
  a nonzero conic constant).
\end{abstract}


\maketitle

\section{Introduction}
\label{intro}

Ballistic electron optics relies on the analogy between the Schr\"odinger
equation for electrons and the Helmholtz equation for classical waves to
construct devices that can image the flow of electrons in high-mobility
semiconductors.\cite{spector:92,houten:95,leroy:03,top03} A variation in
electrostatic potential is analogous to a variation in dielectric constant,
so that a curved gate electrode can have the refractive power of a lens
--- as has been demonstrated in the two-dimensional electron gas of a
GaAs heterostructure.\cite{spector:90,sivan:90} The focal length of this
electrostatic lens depends on its curvature, diverging for a flat electrode.

Focusing of light by a flat lens is possible in media with a negative index
of refraction. This so-called Veselago lens \cite{veselago:60,pendry:04} has
a focal length proportional to the distance between lens and source, rather
than fixed by the lens itself. It is also not limited by the single optical
axis of a curved lens and can have a much wider aperture. Photonic crystals
can provide the negative refraction needed for a flat lens,\cite{notomi:00}
as demonstrated experimentally.\cite{parimi:03,cubukcu:03}

The electronic analog of a Veselago lens was proposed in the context of
graphene,\cite{cheianov:07} based on the negative refraction of
an electron crossing from the conduction band into the valence band. Such
interband crossing requires a \textit{p-n} junction, which is highly
resistive if the interface extends over more than an electron wave
length.\cite{katsnelson:06,cheianov:06} It would be desirable to have
a method for producing a flat lens entirely within the conduction band,
in order to avoid a resistive interface. It is the purpose of this work
to propose such a method, in the context of topological insulators.

Topological insulators have a conducting surface with a Dirac
cone of massless, helical low-energy excitations, reminiscent of
graphene.\cite{Qi10,Has10} Indeed, scanning tunneling microscopy has shown
that backscattering of the surface electrons is inhibited, as expected from
conservation of helicity.\cite{roushan:09,zhang:09b} While the large band
gap topological insulator Bi$_2$Se$_3$ has a nearly circular Dirac cone,
in the smaller band gap material Bi$_2$Te$_3$ the cone is warped in an
hexagonal snowflake-like shape.\cite{zhang09,xia:09,hsieh:09,chen:09,Has09}
The hexagonal warping of the Fermi surface enhances the quantum interference
(Friedel) oscillations in the density of states near an impurity or potential
step,\cite{alpichshev:10,zhou:09,lee:09,wang:10} which for a circular
Fermi surface would be suppressed by conservation of helicity.\cite{fu:09}

The electron focusing considered here is an altogether different,
semiclassical consequence of the hexagonal warping. The flat lens is
formed by a potential step on the surface of the topological insulator,
sufficiently high to change the curvature of the Fermi surface from convex
to concave. The sign change of the curvature leads to negative refraction
and focusing, qualitatively similar to the optical Veselago lens --- but quantitatively different because of the nonuniformity of the curvature (quantified by a nonzero conic constant).

In the following two sections, we derive the negative refraction and the line
of focal points (caustics), as well as the diffraction pattern produced
by a point source. We calculate the curvature and conic constant for the specific case of Bi$_2$Te$_3$. 
We conclude in Sec.\ \ref{discuss} by comparing with the flat lens formed
by a \textit{p-n} junction in graphene\cite{cheianov:07} and by discussing
possible experimental realizations in topological insulators.

\section{Negative refraction at a potential step}
\label{refraction}

\subsection{Negative refraction}
\label{negativeref}

\begin{figure}[tb]
\centerline{\includegraphics[width=0.8\linewidth]{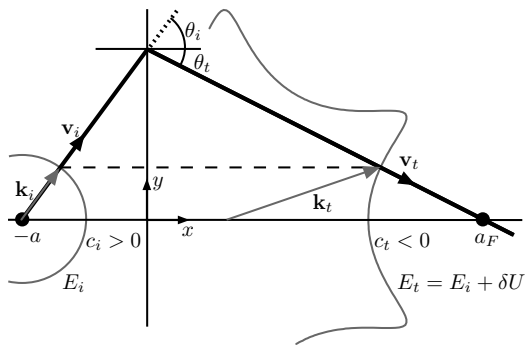}}
  \caption{%
  Negative refraction at a potential step (height $\delta U$) where the
  curvature of the equi-energy contours (thin curves) changes sign from
  $c_{i}>0$ to $c_{t}<0$. An electron (thick arrow) with kinetic energy
  $E_{i}$ is incident at angle $\theta_{i}$ and transmitted at angle
  $\theta_{t}$. Because the curvature changes sign, the electron is
  negatively refracted with $\theta_{t} <0$ for $\theta_{i}>0$.
  }\label{fig:setup}
\end{figure}

Consider an electron propagating approximately along the $x$-axis (the
optical axis) and impinging at $x=0$ onto an electrostatic potential step
$\delta U$ produced by a gate electrode (see Fig.~\ref{fig:setup}). For
simplicity, we assume that the optical axis is parallel to an axis of
crystallographic symmetry, such that the equi-energy contours are $\pm
k_{y}$ symmetric. (For the more general case, see App.\ \ref{tilted}.)
At constant Fermi energy, the kinetic energy changes from $E_{i}$ in the
incident (left) region to $E_{t} = E_{i} + \delta U$ in the transmitted
(right) region. The equi-energy contour at the left is given locally by
$\delta k_{{i},x} = -\frac{1}{2}c_{i} k_{{i},y}^2$, for a two-dimensional
wave vector ${\bm k}_{i}= (k_{{i},0} + \delta k_{{i},x} , k_{{i},y})$
approximately along the optical axis, and similarly $\delta k_{{t},x}
= -\frac{1}{2}c_{t} k_{{t},y}^2$ at the right. The coefficients $c_{i}$
and $c_{t}$ are the curvatures of the Fermi surface for normal incidence,
at the two sides of the potential step.
 
The velocity ${\bm v} = \hbar^{-1}\partial E/\partial{\bm k}$ is normal
to the equi-energy contours, so that the velocities ${\bm v}_{i}$ and
${\bm v}_{t}$ in the left and right regions make, respectively, an angle
$\theta_{i}=c_{i} k_{{i},y}$ and $\theta_{t}= c_{t} k_{{t},y}$ with the
$x$-axis.  Conservation of transverse momentum ($k_{{i},y} = k_{{t},y}$)
leads to the linearized Snell's law
\begin{equation}\label{eq:snel}
\theta_{t}=(c_{t}/c_{i})\theta_{i}, \qquad \text{for }
  \theta_{i},\theta_{t} \ll 1 .
\end{equation}
The inverse curvature plays the role of the refractive index in
optics. Negative refraction (meaning $\theta_{i}\theta_{t}<0$) takes place
when $c_{i}$ and $c_{t}$ have opposite signs, as illustrated in Fig.\
\ref{fig:setup}.

\subsection{Noncircular Snell's law}
\label{Snellslaw}

As we will see in the next section, to calculate the image of a point
source we will need to include the first nonlinear correction to Eq.\
\eqref{eq:snel}. In optics, where one has a circular equi-energy contour,
Snell's law $c_{i}\sin\theta_{t}=c_{t}\sin\theta_{i}$ implies the series
expansion
\begin{equation}
\theta_{t}=n_{1}\theta_{i}+n_{3}\theta_{i}^{3}+{\cal O}(\theta_{i}^{5}),\label{thetaexp}
\end{equation}
with $n_{1}=c_{t}/c_{i}$ and $n_{3}=\frac{1}{6}n_{1}(n_{1}^{2}-1)$. More
generally, we can write
\begin{align}
n_{1}=c_{t}/c_{i},\;\;n_{3}=\tfrac{1}{6}n_{1}(n_{1}^{2}-1)+\Delta,\label{n1n3}
\end{align}
where $\Delta$ quantifies the deviation from the optical Snell's
law.\cite{note1}

The parameter $\Delta$ vanishes for a circular Fermi surface as in
graphene,\cite{cheianov:07,cserti:07,cse09} but is nonzero for the warped
Fermi surfaces of topological insulators. In order to relate $\Delta$ to
the Fermi surface, we parameterize the equi-energy contour using polar
coordinates by ${\bm k}=\kappa(\phi)(\cos\phi,\sin\phi)$, where $\phi$
is the angle between the wave vector ${\bm k}$ and the $x$-axis and
$\kappa=|\bm{k}|$. A subscript $i$ or $t$ distinguishes the parameters at
the two sides of the potential step.

The noncircular Snell's law is expressed by the three equations
\begin{align}
&\kappa_{t}(\phi_{t})\sin\phi_{t}=\kappa_{i}(\phi_{i})\sin\phi_{i},\label{snell1}\\
&\tan\theta_{i}=\frac{\kappa_{i}(\phi_{i})\tan\phi_{i}-\kappa'_{i}(\phi_{i})}{\kappa_{i}(\phi_{i})+\kappa'_{i}(\phi_{i})\tan\phi_{i}},\label{snell2}\\
&\tan\theta_{t}=\frac{\kappa_{t}(\phi_{t})\tan\phi_{t}-\kappa'_{t}(\phi_{t})}{\kappa_{t}(\phi_{t})+\kappa'_{t}(\phi_{t})\tan\phi_{t}},\label{snell3}
\end{align}
where $\kappa'=d\kappa/d\phi$. The first equation expresses the
continuity of the $y$-component of the wave vector at the interface
$x=0$, while the second and third equations relate the angles $\theta$
and $\phi$ of velocity and wave vector (using the fact that ${\bm v}$
is perpendicular to the equi-energy contour). The circular Snell's law
$\kappa_{t}\sin\theta_{t}=\kappa_{i}\sin\theta_{i}$ is recovered for
$\kappa'=0$, when $\theta=\phi$.

Near $\phi=0$, the equi-energy contour can be parameterized in terms of
the curvature $c$ and conic constant $\mathcal{K}$,
\begin{equation}
  k_{y}^{2}=-(2/c)\delta k_{x} -(1+\mathcal{K}) 
  (\delta k_{x})^{2},\label{conicalparam}
\end{equation}
with $\delta k_{x}=k_{x}-\kappa(0)$. The noncircular Snell's law then
expands to Eqs.\ \eqref{thetaexp}--\eqref{n1n3} with
\begin{equation}
  \Delta=\frac{c_{t}}{2c_{i}^{3}}\bigl(c_{t}^{2} \mathcal{K}_{t}-c_{i}^{2}
  \mathcal{K}_{i}\bigr).\label{Deltaresult}
\end{equation}

\subsection{Application to Bi$_{\bm 2}$Te$_{\bm 3}$}
\label{applicationBi}

We apply these general considerations to the topological insulator
Bi$_2$Te$_3$. On the $[111]$ surface and close to the center of the Brillouin
zone (the $\Gamma$ point) the Hamiltonian can be approximated by\cite{fu:09}
\begin{equation}\label{eq:ham}
  H= \hbar v k \bigl(\sigma_y  \cos \phi 
  - \sigma_x\sin \phi   
  + \lambda^2 k^2  \sigma_z\cos 3\phi \bigr).
\end{equation}
The dispersion relation in the conduction band ($E>0$) is
\begin{align}\label{eq:disp}
  E ({\bm k}) &= 
  \hbar v
  \sqrt{ k^2 + \smash{(\lambda^2 k^3 \cos 3 \phi)^2} }  
  \nonumber\\
  &=  \hbar v
  \sqrt{k^2 + \smash{\lambda^4 (k_x^3 - 3 k_x k_y^2)^2}}.
\end{align}
The $\sigma_{i}$'s are Pauli matrices acting on the electron spin and
$\phi$ denotes the angle of the wave vector ${\bm k}$ with respect
to the $\Gamma$K direction in the Brillouin zone (oriented along the
$x$-axis).  The parameters $v \approx 4\cdot 10^{5}\,{\rm m/s}$ and
$\lambda \approx 1\,$nm were estimated by fitting to data from angularly
resolved spectroscopy.\cite{chen:09,hsieh:09} (Additional terms quadratic
in momentum can be included in the fit, but these do not qualitatively
change the dispersion.)

\begin{figure}[tb]
\centerline{\includegraphics[width=0.8\linewidth]{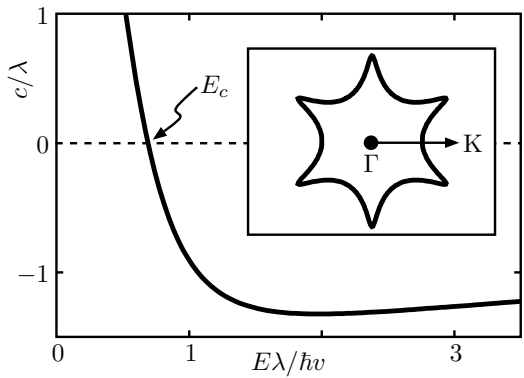}}
  \caption{%
  Curvature $c(E)$ of the equi-energy contour in the $\Gamma$K direction,
  calculated from Eq.\ \eqref{eq:curv}. The shape changes from convex
  to concave at energy $E_{c}$. The maximally negative curvature is $c
  \approx -1.3\,\lambda$ for $E \approx 2\, \hbar v/\lambda$, where the
  equi-energy contour has the snowflake-like shape shown in the inset.
  }\label{fig:curv}
\end{figure}
\begin{figure}[tb]
  \centering
  \includegraphics{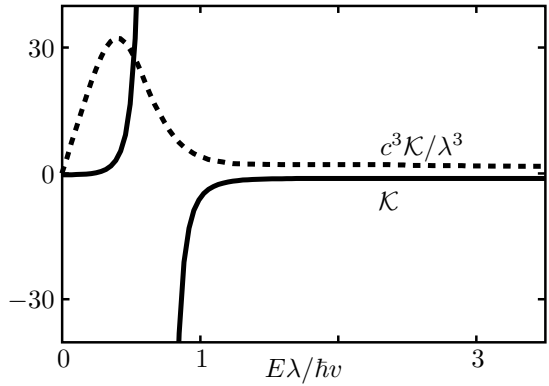}
  \caption{%
  Plot of the conic constant $\mathcal{K}$ (solid line) as well as the
  combination $c^3 \mathcal{K} / \lambda^3$ [appearing in the noncircular
  Snell's law \eqref{Deltaresult}] (dashed line), both as a function of
  the energy $E$. The divergence of $\mathcal{K}$ is at the energy $E_c$
  where the curvature vanishes.
  }\label{fig:conic}
\end{figure}

The curvature $c(E)$ of the equi-energy contour in the $\Gamma$K direction
is given by
\begin{equation}\label{eq:curv}
  c(E) = \lambda \frac{1-6 \varkappa^4}{ \varkappa + 3 \varkappa^5},
\end{equation}
with $\varkappa^2 = \xi_+^{1/3} - \xi_-^{1/3}$ defined in terms of
\begin{equation}\label{eq:xi}
  \xi_\pm = \frac{1}{6 \hbar^2 v^2} \biggl(
  \sqrt{\tfrac{4}{3} \hbar^4 v^4 +9 \lambda^4 E^4} \pm 3 
  \lambda^2 E^2 \biggr) .
\end{equation}
The quantity $\varkappa/\lambda=\kappa(0)$ equals $|\bm{k}|$ at $\phi=0$.

The energy dependence of the curvature is plotted in Fig.\
\ref{fig:curv}. As discovered by Fu,\cite{fu:09} the curvature changes
sign when $\varkappa_{c}^4 = 1/6$, which corresponds to an energy $E_{c}
= 6^{-3/4}\sqrt{7}\, \hbar v/\lambda \approx 0.2\,$eV and a wave vector
$k_{c} = \varkappa_{c}/\lambda\approx 0.6\,{\rm nm}^{-1}$. At the same
point the conic constant
\begin{equation}\label{eq:conic}
  \mathcal{K}(E) =  \frac{3 \varkappa^4 (35 - 60 \varkappa^4 + 72 \varkappa^8)}{(1-6 \varkappa^4)^3}
 \end{equation}
diverges and thereby changes sign, cf.\ Fig.~\ref{fig:conic}.

\section{Caustics from a point source}
\label{caustics}

\subsection{Focusing of classical trajectories}
\label{trajectories}

\begin{figure}[tb]
\centerline{\includegraphics[width=0.8\linewidth]{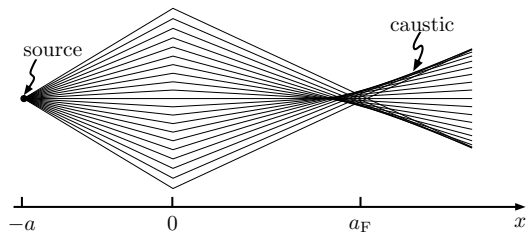}}
  \caption{%
  Classical trajectories refracted at a potential step at $x=0$, with the
  cusp caustic indicated.
  }\label{fig:caustic_plot}
\end{figure}

Because of the negative refraction, diverging trajectories become
converging at the potential step and then cross at a focal point (see
Fig.\ \ref{fig:caustic_plot}). If a point source is placed at $(-a,0)$,
a distance $a$ from the interface at $x=0$, then the trajectory for an
electron incident at an angle $\theta_{i}$ and transmitted at an angle
$\theta_{t}$ is parameterized by
\begin{equation}\label{eq:traj}
  y(x; \theta_{i}) = 
  \begin{cases}
    (a+x) \tan \theta_{i}, & {\rm for}\;\; x<0, \\
    a \tan \theta_{i} + x \tan \theta_{t}, & {\rm for}\;\; x>0.
  \end{cases}
\end{equation}
On the optical axis $y,\theta_{i},\theta_{t}\rightarrow 0$ we obtain the
focal point $(a_{F},0)$, with
\begin{equation}\label{eq:focus}
  a_{F} = -a/n_{1}=-\frac{c_{i}}{c_{t}} a,
\end{equation}
proportional to the ratio of the two curvatures. As in the optical Veselago
lens,\cite{Any08,She08} the focal point is displaced from the optical
axis as we increase the angle of incidence, so that the point $(a_{F},0)$
is the cusp on a curve of focal points. This caustic curve (called an
\textit{astroid}\cite{Loc61}) is visible in Fig.\ \ref{fig:caustic_plot}
as the envelope of the refracted trajectories.

The caustic curve near $(a_{F},0)$ is obtained from Eq.\ \eqref{eq:traj}
and the nonlinear Snell's law \eqref{thetaexp} by demanding that $\partial
y/\partial\theta_{i}=0$. We find
\begin{equation}
\alpha(y/a)^{2}=(x/a_{F}-1)^{3},\label{astroid}
\end{equation}
with the opening rate of the cusp governed by the parameter
\begin{equation}
\alpha=\frac{27}{8}\bigl[1-(c_{t}/c_{i})^{2}-2(c_{i}/c_{t})\Delta\bigr].
\label{alphadef}
\end{equation}
For $\Delta=0$, so for a circular Fermi surface, this agrees with Refs.\
\onlinecite{cheianov:07,She08}. Depending on the sign of $\alpha$, the
cusp points away from the potential step (for $\alpha >0$) or towards the
potential step (for $\alpha<0$). For $\alpha=0$ higher than third-order
terms in the expansion \eqref{thetaexp} have to be included in order to
obtain the caustic curve.

\subsection{Quantum interference near the focal point}
\label{interference}

\begin{figure}[tb]
  \centering
  \includegraphics[width=0.8\linewidth]{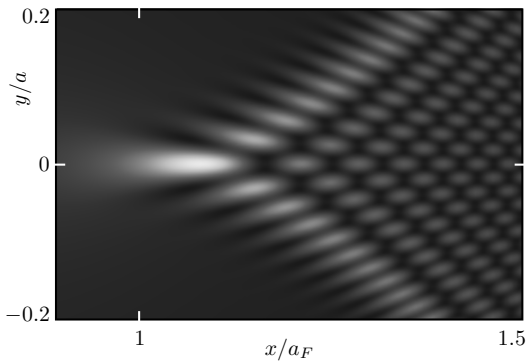}
  \caption{%
  Gray-scale plot of the current density $j(\mathbf{r})$ as a function
  of position $\mathbf{r}= (x,y)$ near the focal point $(a_F,0)$ for a
  source located at $(-a,0)$, calculated from Eqs.\ \eqref{Phiresult}
  and \eqref{jresult} for $\alpha=1$ and $a/c_i = 100$. Lighter shades of
  gray indicate higher current densities. The cusp caustic starting at the
  focal point is decorated by oscillations on the scale of the wavelength.
  }\label{fig:dens}
\end{figure}

The diffraction pattern near a cusp caustic has a universal functional form
(Pearcey integral),\cite{Pea46,Ber80} but the parameters governing that
function are modified for noncircular equi-energy contours. We calculate the
wave function $\Psi$ at a point $\bm{r}=(x,y)$ near the cusp by summing
over partial waves $\Psi_{y_{0}}$ from points $\bm{r}_{0}=(0,y_{0})$
along the potential step [excited by a point source at $\bm{r}_{\rm
source}=(-a,0)$]. In the far-field approximation, for $a$ and $a_F$ large
compared to the wave length, the partial waves have the simple form
\begin{align}
&\Psi_{y_{0}}=\begin{pmatrix}
u_{y_{0}}\\ v_{y_{0}}
\end{pmatrix}A_{y_{0}}e^{i\Phi_{y_{0}}},\label{Psidef}\\
&\Phi_{y_{0}}=\bm{k_{i}}\cdot(\bm{r}_{0}-\bm{r}_{\rm source})+\bm{k}_{t}\cdot(\bm{r}-\bm{r}_{0}).\label{Phidef}
\end{align}
The amplitude $A_{y_{0}}$ and spinor components $u_{y_{0}},v_{y_{0}}$
vary slowly as $y_{0}$ is varied on the scale of the wave length, so
we fix their values at $A_{0},u_{0},v_{0}$ and retain only the $y_{0}$
dependence of the phase $\Phi_{y_{0}}$.

In the optical case, the wave vectors $\bm{k}_{i}$ and $\bm{k}_{t}$ at
the two sides of the interface point in the direction of the velocity
and hence are parallel to the rays $\bm{r}_{0}-\bm{r}_\text{source}$
and $\bm{r}-\bm{r}_{0}$. For a noncircular Fermi surface this is no
longer true and we have to take into account the difference between the
angles $\phi_{i}$, $\phi_{t}$ and $\theta_{i}={\rm arctan}\,(y_{0}/a)$
and $\theta_{t}={\rm arctan}\,[(y-y_{0})/x]$ which the wave vectors and
the rays make with the $x$-axis. The relation between $\phi$ and $\theta$
is expressed by Eqs.\ \eqref{snell2} and \eqref{snell3}, in terms of the
radial parameter $\kappa(\phi)=|\bm{k}|$ of the equi-energy contour.

We expand $\Phi_{y_{0}}$ in a power series in $y_{0}$. Near the cusp caustic
\eqref{astroid}, $y/a={\cal O}(y_{0}/a)^{3}$ while $x/a_{F}-1={\cal
O}(y_{0}/a_{F})^{2}$. To fourth order in $y_{0}$ we find
\begin{multline}\label{Phiresult}
  \Phi_{y_0} =  \kappa_{i}(0)a+\kappa_{t}(0)x \\
   -\frac{y y_0}{c_t a_F} -
   \frac{ (x - a_F) y_0^2}{2 c_t a_F^2} + \frac{\alpha y_0^4}{27 c_t a_F a^2},
\end{multline}
with $\alpha$ given by Eq.\ \eqref{alphadef}. One readily checks
that the stationary phase equations $\partial\Phi_{y_{0}}/\partial
y_{0}=0=\partial^{2}\Phi_{y_{0}}/\partial y_{0}^{2}$ give the caustic curve
\eqref{astroid}. (These equations correspond to the geometric optics limit
$c_t \to 0$ of vanishing wavelength.)

The current density $j(\mathbf{r})$ follows upon integration over $y_{0}$,
\begin{equation}
j(\mathbf{r})=j_{0}\left|\int_{-\infty}^{\infty}
dy_{0}\,e^{i\Phi_{y_{0}}}\right|^{2},\label{jresult}
\end{equation}
with $j_{0}$ a constant proportional to the product of the injection
rate at the source and the transmission probability $T$ through the
potential step. By rescaling the integration variable $y_0 \to a y'_0$,
we see that the current density \eqref{jresult} as a function of $x/a_F$
and $y/a$ depends only on the two parameters $\alpha$ and $a/c_i$. Figure
\ref{fig:dens} is a plot of this current density, showing the characteristic
interference pattern of a cusp caustic.

\subsection{Focusing by a flat lens}
\label{fotcusing}

\begin{figure}[tb]
 \centerline{\includegraphics[width=0.8\linewidth]{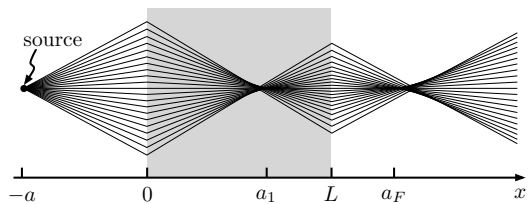}}
  \caption{%
  Flat lens, with two potential steps (upward at $x=0$ and downward at
  $x=L$) and two cusp caustics (at $x=a_{1}$ and $x=a_{F}$).
  }\label{fig:flatlens}
\end{figure}

The flat lens in Fig.\ \ref{fig:flatlens} is formed by the potential
profile $U(x)=\delta U$ for $0<x<L$, $U(x)=0$ otherwise. We denote the
Fermi surface curvatures (of opposite sign) inside the lens ($0<x<L$) by
$c_{\rm lens}$ and outside ($x<0$, $x>L$) by $c_{0}$. Negative refraction
at the two potential steps at $x=0$ and $x=L$ focuses a source at $x=-a$
on the optical axis ($y=0$) first onto the point $a_{1}=-(c_{0}/c_{\rm
lens})a$ inside the lens and then onto the point
\begin{equation}
a_{F}=(1-c_{\rm lens}/c_{0})L-a\label{aF2}
\end{equation}
outside the lens (provided it is sufficiently thick, $|c_{\rm
lens}L|>|c_{0}a|$).

The classical trajectories are now parameterized by
\begin{equation}\label{eq:tra2j}
  y(x; \theta_{i}) = 
  \begin{cases}
    (a+x) \tan \theta_{i}, \;\; {\rm for}\;\; x<0, \\
    a \tan \theta_{i} + x \tan \theta_{t}, \;\; {\rm for}\;\; 0<x<L,\\
    (a+x-L) \tan \theta_{i} + L \tan \theta_{t},\;\; {\rm for}\;\; x>L.
  \end{cases}
\end{equation}
The relation between $\theta_{t}$ and $\theta_{i}$ is still given by Eq.\
\eqref{thetaexp}, with
\begin{align}
n_{1}=c_{\rm lens}/c_{0},\;\;n_{3}=\tfrac{1}{6}n_{1}(n_{1}^{2}-1)+\Delta.\label{n1n3lens}
\end{align}
The cusp caustic near $(a_{F},0)$ has the form
\begin{equation}
\beta(y/a)^{2}=(x/a_{F}-1)^{3},\label{astroid2}
\end{equation}
as in Eq.\ \eqref{astroid} but with a different parameter
\begin{equation}
\beta=\frac{27}{8}\frac{La^{2}}{a_{F}^{3}}\frac{c_{\rm lens}}{c_{0}}\bigl[1-(c_{\rm lens}/c_{0})^{2}-2(c_{0}/c_{\rm lens})\Delta\bigr].
\label{betadef}
\end{equation}
Notice that $\alpha$ and $\beta$ have the opposite sign (because of the
factor $c_{\rm lens}/c_{0}<0$), so that the cusps inside and outside the
lens point in opposite directions (as visible in Fig.\ \ref{fig:flatlens}).

The flat-lens diffraction pattern near the caustic is given by the same
Pearcey integral \eqref{Phiresult}--\eqref{jresult} as for a single
interface, but with different coefficients,
\begin{multline}
j(\bm{r})= j_{0}\biggl|\int_{-\infty}^{\infty}
 d y_{0}\,
 \exp\biggl[-\frac{yy_{0}}{c_{0}a_{F}}-\frac{(x-a_{F})y_{0}^{2}}{2c_{0}a_{F}^{2}}\\
+\frac{\beta y_{0}^{4}}{27\,c_{0}a_{F}a^{2}}\biggr]\biggr|^{2}.\label{j2result}
\end{multline}
Thus, the interference pattern that can be observed near $a_F$ looks
similar to Fig.~\ref{fig:dens}.

\section{Discussion}
\label{discuss}

\subsection{Intraband versus interband negative refraction}
\label{comparison}

The Veselago lens at a \textit{p-n} junction in graphene\cite{cheianov:07} uses \textit{interband} scattering to achieve negative refraction. In contrast, the mechanism considered here is \textit{intraband}, operating entirely within the conduction band. The \textit{p-n} junction has one special feature which our setup lacks, which is the possibility to use electron-hole symmetry to collapse the caustic curve onto a single focal point (when $c_{t}=-c_{i}$). In our setup the Fermi surfaces at the two sides of the potential step are not related by any symmetry relation, so in general the two Fermi surface curvatures $c_{t}$ and $c_{i}$ will be different in magnitude.

The main advantage of an intraband over an interband mechanism for negative refraction is that the transmission probability $T$ can be much higher. Typically, the width $d$ of the potential step will be large compared to the Fermi wave length $\lambda_{F}=2\pi/k_{F}$. Intraband transmission is then realized with unit probability, up to exponentially small backscattering corrections: $T=1-{\cal O}(e^{-k_{F}d})$. Interband transmission, in contrast, has $T\simeq\exp(-k_{F}d\sin^{2}\theta_{i})$, so it is exponentially suppressed for angles further than $\sqrt{\lambda_{F}/d}$ from normal incidence.\cite{cheianov:06}

\subsection{Experimental realization}
\label{realization}

Realization of the intraband flat lens proposed here, requires firstly a topological
insulator with sufficiently long mean free paths to ensure
ballistic motion of the electrons from source to focus. Sufficiently pure single crystals should make this possible. 

Secondly, and more specifically, the curvature of the Fermi surface should be tunable from positive to negative values by a gate voltage. From spectroscopic data\cite{chen:09} for Sn-doped Bi$_2$Te$_3$ we would estimate that a potential step $\delta U \simeq -0.1 \,$eV would produce a positive curvature inside a narrow strip and a negative curvature outside (as in Fig.~\ref{fig:flatlens}). The strip itself would also allow for bulk conduction, because in Bi$_2$Te$_3$ a positively curved Dirac cone of surface states overlaps with bulk states. Since the regions outside the lens have only surface conduction, we do not expect the bulk states inside the lens to spoil the focusing.  

A point source can be created, for example, using the ``needle-anvil'' technique developed for point contact spectroscopy,\cite{Yanson} or alternatively using a scanning tunneling microscope (STM). For the spatially resolved detection of the current density distribution an STM tip is most convenient. Such a setup would provide a sensitive probe of the nonspherical Fermi surface of a topological insulator, in a similar way as has recently been proposed for metals.\cite{weismann:09}

\acknowledgments

We acknowledge fruitful discussions with M.\ Wimmer. This research was
supported by the Dutch Science Foundation NWO/FOM and by an ERC Advanced
Investigator grant.

\appendix

\section{Sheared caustic curve for tilted potential step}
\label{tilted}

In the main text we have assumed for simplicity that the potential step is
perpendicular to the $\Gamma$K direction in Fig.\ \ref{fig:curv}. Then only
odd powers of $\theta_{i}$ appear in the expansion \eqref{thetaexp}. If
the potential step is tilted relative to this crystallographic axis,
then the cusp caustic persists but in a distorted form, as we now derive.

Including also even powers of $\theta_{i}$ in Eq.\ \eqref{thetaexp} one
would have the expansion
\begin{equation}
\theta_{t}=n'_{0}+n'_{1}\theta_{i}+n'_{2}\theta_{i}^{2}
+n'_{3}\theta_{i}^{3}+{\cal O}(\theta_{i}^{4}).\label{thetaexp2} 
\end{equation} 
By rotating the coordinate axis, we can set $n'_{0}=0$. The expressions
simplify if we expand in powers of $\tan\theta_{i}$,
\begin{equation}
 \tan\theta_t =
 m_{1}\tan\theta_i+m_{2}\tan^2\theta_i+m_{3}\tan^3\theta_i+{\cal
 O}(\tan^{4}\theta_{i}).
\end{equation}
From Eq.\ \eqref{eq:traj}, demanding $\partial y/\partial\theta_{i}=0$,
we obtain the implicit caustic equation
\begin{align}\label{implicit} \begin{pmatrix}
  x\\y
\end{pmatrix}={}&a({m_{1}+ 2
m_{2}\tan\theta_i+3m_{3}\tan^2\theta_i})^{-1}\nonumber\\
&\times\begin{pmatrix} -1\\ (m_{2}+2m_{3}\tan\theta_i)\tan^2\theta_i
\end{pmatrix}.  
\end{align} 
The cusp of the caustic is given by the condition $\partial
x/\partial\theta_i=0$. It is at $\tan\theta_{i0}=-m_{2}/3m_{3}$. In order
to remain in the region of validity of the expansion \eqref{thetaexp2},
we assume that $|m_2| \ll |m_3|$ so that the tilt remains small. Then the
cusp is located at
\begin{equation}
 \begin{pmatrix}
  x_0\\y_0
 \end{pmatrix}=
\begin{pmatrix} -a/m_1\\ 0 \end{pmatrix}.  \end{equation}

We now expand Eq.~\eqref{implicit} near $\tan\theta_i=\tan\theta_{i0}$
to third order in $\delta=\tan\theta_i-\tan\theta_{i0}$,
\begin{equation}
\begin{pmatrix}
  x-x_0\\y-y_0
\end{pmatrix}= a m_3 m_1^{-1} \begin{pmatrix}
3 m_1\delta^2\\
-m_2 m_3^{-1} \delta^2 +2\delta^3
\end{pmatrix}.
\end{equation}
Eliminating $\delta$ yields the caustic curve
\begin{equation}
 \gamma [ y-y_0+ \epsilon (x-x_0) ]^2=(x-x_0)^3,\label{sheared}
\end{equation}
with coefficients $\gamma = 27 a m_3/4 m_1^4$ and $\epsilon= m_1 m_2 /3 m_3$.
Equation \eqref{sheared} has the general form of a sheared cusp caustic
from catastrophe theory.\cite{Nye84}


\begin{thebibliography}{10}
\bibitem{note1}
We assume that the potential step is perpendicular to the $\Gamma$K direction
in Fig.\ \ref{fig:curv}, so that only odd powers of $\theta_{i}$ appear in the expansion \eqref{thetaexp}. This assumption is relaxed in App.\ \ref{tilted}.
\end{thebibliography}
\end{document}